\documentclass{optica-article}
\journal{opticajournal} 
\articletype{Research Article}

\usepackage{lineno}
\usepackage{subcaption}
\usepackage{graphicx}
\usepackage[nolist]{acronym}

\begin{acronym}[DMWG] 
    \acro{cw}{continuous wave}
    \acro{SNR}{signal-to-noise ratio}
    \acro{RF}{radio frequency}
    \acro{FWHM}{full width at half maximum}
    \acro{DMWG}{double-metal waveguide}
    \acro{QCL}{Quantum Cascade Laser}
    \acro{FSR}{free spectral range}
    \acro{GVD}{Group Velocity Dispersion}
    \acro{THz QCLs}{Terahertz Quantum Cascade Lasers}
    \acro{THz}{terahertz}
    \acro{QC}{quantum cascade}
    \acro{FTIR}{Fourier transform infrared spectrometer}
    \acro{NDR}{negative differential resistance}
    \acro{LIV}{light-current-voltage}
    \acro{FWM}{four-wave mixing}
\end{acronym}

\begin{document}

\title{Symmetry Control in Coupled Quantum Cascade Ring Lasers}


\author{Dominik Schock,\authormark{1,2,*} Anna Invernici,\authormark{1,2} Michael Jaidl,\authormark{1,2} Miriam Giparakis,\authormark{1,2} 
Sascha Preu,\authormark{4} Aaron M. Andrews,\authormark{2,3} Juraj Darmo,\authormark{1,2} and Karl Unterrainer\authormark{1,2}}

\address{
\authormark{1}Photonics Institute, TU Wien, Gusshausstrasse 27-29, 1040 Vienna, Austria\\
\authormark{2}Centre for Micro- and Nanostructures, TU Wien, Gusshausstrasse 25a, 1040 Vienna, Austria\\
\authormark{3}Institute of Solid-State Electronics, TU Wien, Gusshausstrasse 25a, 1040 Vienna, Austria\\
\authormark{4}Department of Electrical Engineering and Information Technology, TU Darmstadt, 64283 Darmstadt, Germany
}

\email{\authormark{*}dominik.schock@tuwien.ac.at} 


\begin{abstract*} 
Optical frequency combs with fast tunability are essential for integrated spectroscopy and sensing. However, achieving such functionality in the terahertz regime presents a significant challenge. This work introduces a novel approach that exploits supermode symmetry control in terahertz lasers. By electrically manipulating the gain ratio between coupled concentric ring lasers, we selectively excite symmetric and antisymmetric supermodes, allowing frequency tuning and spectral modulation across different frequency regimes and operation modes. Moreover, simultaneous lasing of both supermodes results in a reduced total integrated dispersion, widening the comb locking range. Supported by matrix formalism and numerical simulations, this work demonstrates the potential of symmetry control for enhanced electro-optic modulation, opening new avenues for on-chip spectroscopy and sensing applications.
\end{abstract*}

\section{Introduction}
\ac{THz QCLs} have emerged as leading candidates for coherent, high-power on-chip \ac{THz} sources due to their compact size and silicon chip integration potential\,\cite{Ertl2026, Williams_07, Muller_25, Jaidl_22}. One particularly promising aspect is the generation of frequency combs, given their wide range of applications\,\cite{Yasui_11, Sterczewski_19, Theiner_21}. Although techniques for generating frequency combs, such as converting femtosecond laser pulses into THz frequency combs using photoconductive antennas\,\cite{Yasui_06} or downconverting near-infrared electro-optic frequency combs via photomixing\,\cite{Theiner_21} are established, these methods face challenges including weak output power, limited tunability, high cost, and bulkiness. Extensive research has led to the generation of frequency combs using THz ridge QCLs by utilizing dispersion compensation\,\cite{Burghoff_14} or a careful design of the active region\,\cite{Rsch_2014}. Ring-shaped resonators have proven especially reliable in facilitating comb generation due to the increased nonlinearities resulting from the high mode confinement\,\cite{Jaidl_21, Udem_02}. Realizing the full potential of \ac{THz QCLs} for high-speed applications remains challenging due to the physical limitations of direct electrical modulation. Recent findings in single-mode \ac{THz} lasers have shown that tuning the gain/loss ratio in coupled lasers can enable optical spectrum modulation and power enhancement\,\cite{Brandstetter2014,Ertl_2024}. In this work we present a novel approach for fast electro-optic modulation of the optical spectrum in \ac{THz QCLs} by using two concentric ring lasers with identical width and height placed at a sub-wavelength distance and biased independently. This leads to the potential formation of symmetric and antisymmetric supermodes which can be modulated by tuning the gain/loss ratio. While similar designs have been previously explored for dispersion compensation using two concentric lasers with a common contact \cite{Micheletti_2023}, this work extends their application to electro-optic modulation in THz QCLs. This enables switching between single-mode and comb emission, as well as tuning of the emission spectra between different frequency regimes ranging from 2.28\,THz to 2.55\,THz.
\section{Theory and Design}
Coupled laser resonators can be described by a non-Hermitian matrix, which neglects gain competition and saturation effects while providing insight into the underlying physical principles \cite{Brandstetter_2014,Liertzer_2012, Haus_1991}: 
\begin{equation}\label{eq:hemiltonian_2x2}
H = \begin{pmatrix} 
\omega_a + i g_a & \kappa \\ 
\kappa & \omega_b + i g_b 
\end{pmatrix}.
\end{equation}
The parameters $\omega_a$ and $\omega_b$ represent the resonance frequencies of the two coupled systems, while $g_a$ and $g_b$ denote their respective gain parameters, and $\kappa$ quantifies the inter-system coupling strength. The gain parameter includes the optical gain (Eq.\,\eqref{eq:susceptibility}) as well as radiative and material losses (Eq.\,\eqref{eq:drude_lorentz}). Each ring resonator supports a set of discrete resonance frequencies, separated by the \ac{FSR}. The \ac{FSR} is itself frequency-dependent, and its value is determined by the complex susceptibility. The gain and frequencies of all uncoupled resonances must be determined via the gain-induced dispersion (Eq.\,\eqref{eq:susceptibility}), while accounting for the different lengths of the two resonators.
\begin{figure}[htbp]
\centering\includegraphics[width=6.6cm]{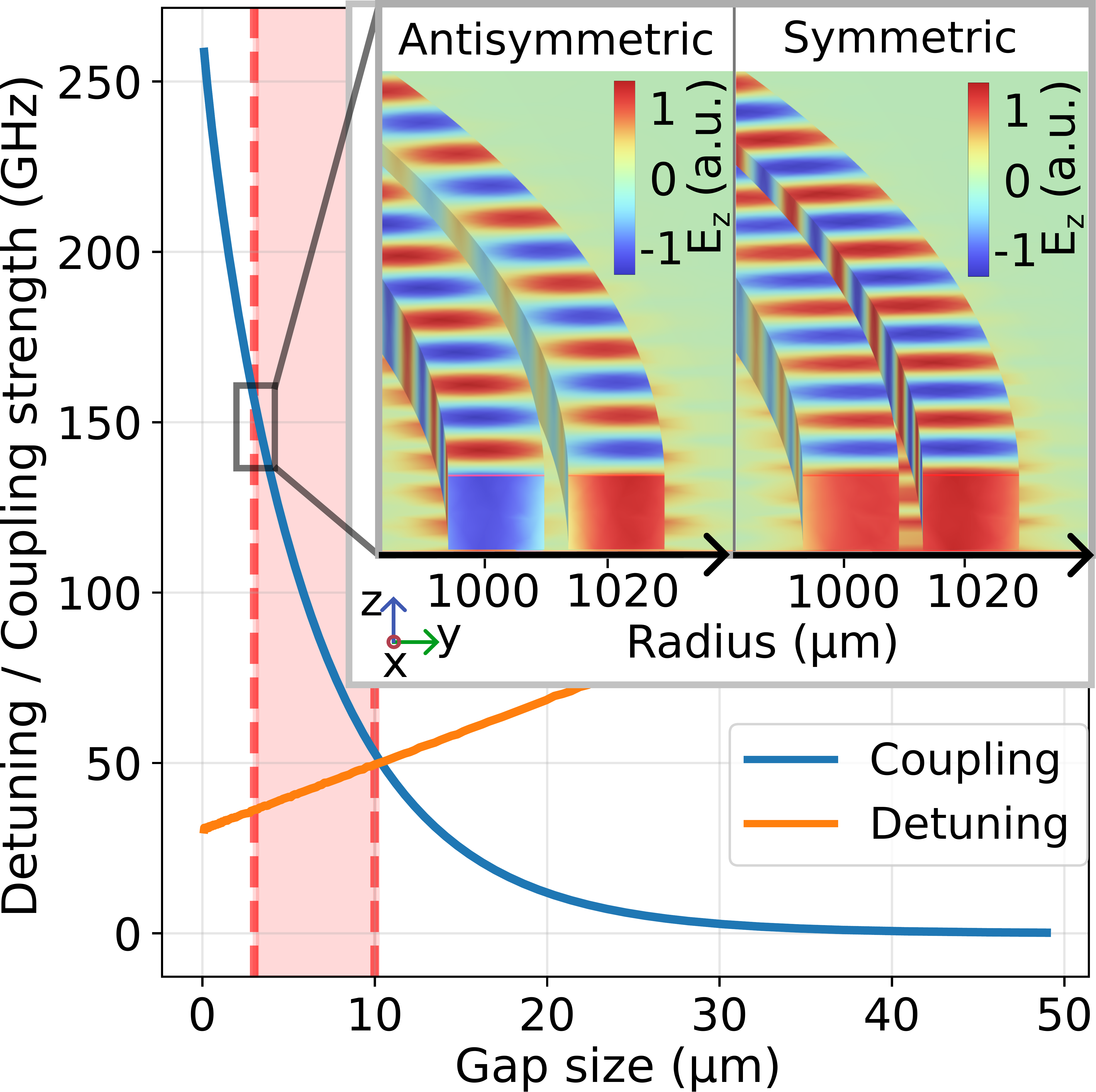}
\caption{Coupling strength and detuning of the lasers depending on the width of the air gap between the lasers. Calculated from the 2D rotational symmetric eigenvalue simulation of the resonator system performed with COMSOL Multiphysics. The coupling strength and detuning was calculated for an identical azimuthal mode     order in both rings. The final designs featured gaps ranging from 4 to 10\,$\mathrm{\mu m}$ as indicated by the red area. For a gap below 10\,$\mu m$ the coupling exceeds the detuning. The \ac{FSR} of the uncoupled system is about 13\,GHz and therefore much smaller than the coupling strength. Inset: Electric field in the z-direction, normal to the metal-semiconductor interface. Simulated for the symmetric and antisymmetric supermodes for a gap size of 4\,$\mu m$.}
\label{fig:e_field_coupling}
\end{figure}
This leads to a Vernier-like distribution of the uncoupled modes. As the coupling strength in our system is more than 10 times larger than the \ac{FSR} of the uncoupled system  ($\sim 13$\,GHz) and much larger than the detuning, multiple resonances are coupled (Fig.\,\ref{fig:e_field_coupling}). Because of the strong coupling the Vernier effect is averaged out, forming symmetric and antisymmetric supermodes. This system can still be solved very conveniently within a $2 \times 2$ matrix formalism by assuming a constant mode spacing in the uncoupled system which allows a transition from a global eigenmode representation to a basis defined by the \ac{FSR} of the two uncoupled resonators (Eq.\,\eqref{eq:hemiltonian_2x2_FSR}). This allows a direct calculation of the \ac{FSR} for different detuning, gain, and coupling conditions.
\begin{equation}\label{eq:hemiltonian_2x2_FSR}
H = \begin{pmatrix} 
\mathrm{FSR}_a + i \tilde{g_a} & \tilde{\kappa} \\ 
\tilde{\kappa} & \mathrm{FSR}_b + i \tilde{g_b} 
\end{pmatrix}
\end{equation}
The \ac{FSR}s of the uncoupled lasers are denoted as $\mathrm{FSR}_a$ and $\mathrm{FSR}_b$. Consequently, the coupling and the gain must be expressed in terms of the round-trip frequency. Thus, $\tilde{g_a}$ and $\tilde{g_b}$ are defined as the gain per round-trip, and $\tilde{\kappa}$ as the coupling coefficient per round-trip.
\begin{figure}[h]
\centering\includegraphics[width=13cm]{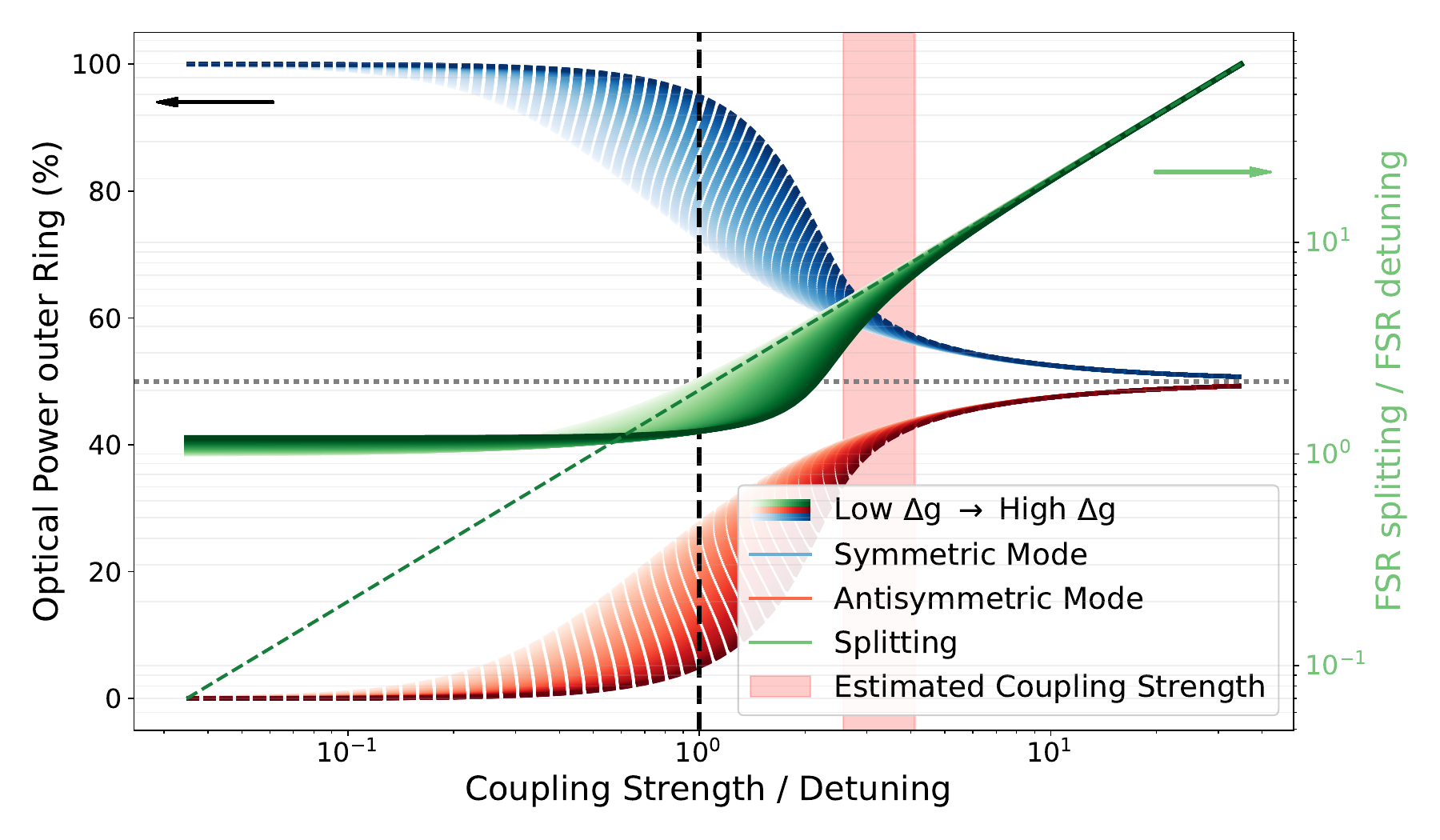}
\caption{Hybridization calculated using the matrix formalism in Eq.\,\eqref{eq:hemiltonian_2x2}, with the gain/loss difference $|\tilde{g_a} - \tilde{g_b}| = \Delta g$ between 0\,cm$^{-1}$\,and\,20\,cm$^{-1}$. The gain/loss difference is depicted as color gradient from bright (low detuning) to dark (high detuning). Higher gain detuning leads to lower \ac{FSR} splitting and lower hybridization. Depending on the coupling strength, the symmetric and antisymmetric modes get hybridized. Even at very high coupling strengths, the symmetric mode is mostly confined to the outer ring, while the antisymmetric mode is predominantly localized in the inner ring.}
\label{fig:hybridisation}
\end{figure}
By calculating the eigenvalues and eigenvectors, one can extract the frequencies and gains of all symmetric and antisymmetric modes. The symmetric modes are predominantly localized in the ring with a lower natural frequency, as the symmetric mode is the lower energy mode. Conversely, the antisymmetric modes are dominantly localized in the ring with higher resonance frequency (Fig.\,\ref{fig:hybridisation}). Given that the round-trip frequency is defined by the length, the symmetric mode will be mainly localized in the outer ring and the antisymmetric mode in the inner ring. The ratio of this distribution depends primarily on the ratio of coupling strength to detuning between the uncoupled \ac{FSR}s. As illustrated in Fig.\,\ref{fig:hybridisation} increasing values of this ratio typically result in a higher degree of hybridization. The coupling strength can be tuned by changing the gap between the lasers (Fig.\,\ref{fig:e_field_coupling}). The width of the gap not only defines the coupling strength but also the detuning between the resonators, as it alters the length difference. For gaps narrower than 10\,$\mu m$ calculated for identical azimuthal mode order in both rings, the coupling strength exceeds the detuning. This is necessary to achieve high hybridization of the supermodes and thereby defines the maximum gap size. Specifically, for a gap size of 4\,$\mu m$, the detuning is approximately 38\,GHz, while the coupling strength is around 135\,GHz, or about 3.55 times larger. This leads to a high degree of hybridization ($P_{\text{outer sym}}/P_{\text{outer anti}}\sim 70\% $). Another critical design parameter is the width of the waveguide.
\begin{figure}[htbp]
\centering\includegraphics[width=6cm]{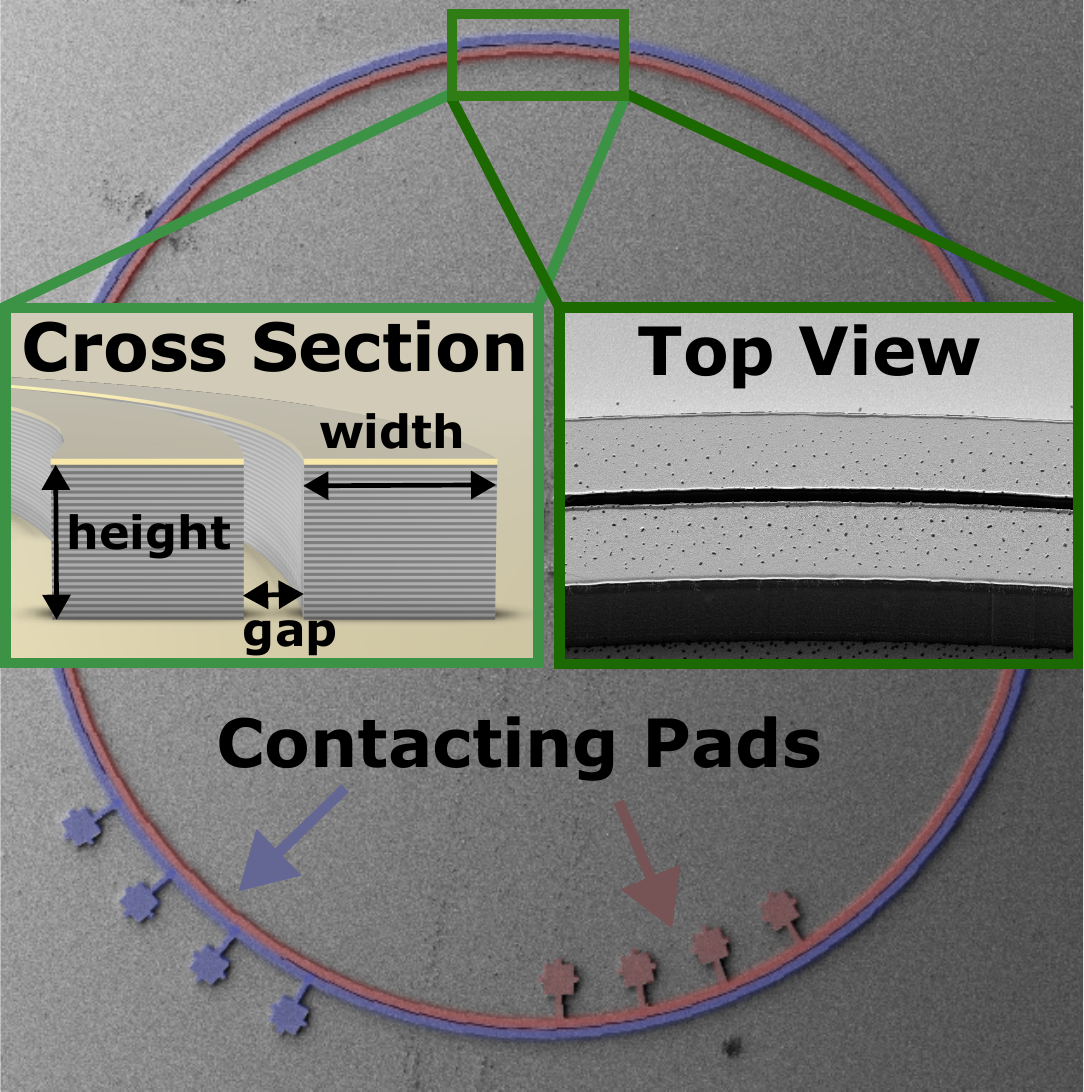}
\caption{SEM image of the concentric ring design including the four contacting pads per laser, enabling independent biasing and \ac{cw} operation. Inset: zoomed-in SEM image and 3D illustration of the cross-sectional view. The key dimensions are as follows: both rings have identical widths of 16\,$\mathrm{\mu m}$ and heights of 13\,$\mathrm{\mu m}$. The inner ring has a radius of 1000\,$\mathrm{\mu m}$, measured at the inside edge, with a gap between the rings ranging from 4\,$\mathrm{\mu m}$ to 10\,$\mathrm{\mu m}$}
\label{fig:C1439_rings_paper}
\end{figure}
In general, double metal waveguides have a very high mode confinement close to unity. Nevertheless, for small width/wavelength ratios the losses increase significantly\,\cite{Williams_2005}. Due to the strong coupling the field distribution in the individual waveguides is not homogeneous and depends on the supermode type. In the case of narrow spacing between the lasers, the overlap integral between the uncoupled waveguide modes has to be taken into account\,\cite{Kapon_1984,Hardy_1985}. Assuming two identical straight waveguides with normalized power, the transversal electric field of the symmetric and antisymmetric mode can be expressed as\,\cite{Hardy_1985}:
\begin{align}
E_t^{\text{sym}} &\propto \frac{1}{\sqrt{1+C}}\left(E_t^{(a)} + E_t^{(b)}\right) \label{eq:sym_mode} \\
E_t^{\text{anti}} &\propto \frac{1}{\sqrt{1-C}}\left(E_t^{(a)} - E_t^{(b)}\right) \label{eq:anti_mode}
\end{align}
with the overlap integral of the uncoupled resonators denoted as $C$. This prefactor leads to an increase in the electrical field in the center of the structure for the symmetric mode, whereas the antisymmetric mode concentrates towards the outer edges of the waveguides (Fig.\,S4). This concentration effect causes an increased evanescent field, for the antisymmetric mode compared to that of the symmetric mode, leading to an increased field amplitude observed on the outer edges of the structure. By exploiting these concentration effects, it is possible to manipulate the losses of each supermode independently. By decreasing the width of the rings, the losses in the symmetric mode remain nearly unchanged, whereas the losses in the antisymmetric mode increase significantly. Decreasing gap size also mostly increases the losses in the antisymmetric mode as for very small gaps the mode gets pushed to the outer edges, increasing the losses, similar to decreasing width. Since the waveguide width dominates the gain/loss ratio without a significant influence on coupling strength and detuning compared to the gap size, the waveguide width serves as primary design parameter for selectively adjusting the gain/loss ratio between the symmetric and the antisymmetric modes. Based on this, an identical width of 16\,$\mu m$ was selected for both rings. This balances the losses of the supermodes, allowing the switching between symmetric and antisymmetric modes by precisely controlling the gain of the inner and outer rings (Fig.\,S5). The gain changes not only the complex part of the susceptibility, but also the real part. Therefore, for the numeric simulation, the different gains of the two lasers were modeled by different complex susceptibilities of the active regions (Eq.\,\eqref{eq:susceptibility}).\\
The complex susceptibility of a \ac{QCL} is determined by the gain region, the waveguide geometry, and the background material properties. The permittivity of the active region material is modeled using a Drude-Lorentz approach, which explicitly accounts for the free-carrier contribution arising from the high doped contact layers\,\cite{Wiley_05}.
\begin{equation}\label{eq:drude_lorentz}
\varepsilon(\omega) = 
\underbrace{\varepsilon_\infty \left[ 1 + \frac{\omega_{LO}^2 - \omega_{TO}^2}{\omega_{TO}^2 - \omega^2 - i \gamma_{PH} \omega} \right]}_{\text{Lorentz contribution}}
- 
\underbrace{\frac{\omega_p^2}{\omega^2 + i \omega \gamma_D}}_{\text{Drude contribution}}
\end{equation}
In Eq.\,\eqref{eq:drude_lorentz} $\varepsilon_\infty$ is the high-frequency dielectric constant, $\omega_{LO}$ and $\omega_{TO}$ are the longitudinal and transverse optical phonon frequencies, $\gamma_{PH}$ is the phonon damping constant, $\omega$ is the angular frequency, $\omega_p$ is the plasma frequency of the free carriers, and $\gamma_D$ is the plasma damping frequency.
Other critical contributions are the losses in the top and bottom gold, which are modeled by experimental data following Ordal 1978 \cite{Ordal_1987}.
The complex susceptibility of the gain region of a \ac{QCL} is described by the following expression \cite{Willenberg_2003, Opacak_2021}:
\begin{equation}\label{eq:susceptibility}
\chi(\omega) =\frac{\omega_0\, g_c \, c_0\, n_r}{\omega^2 \, }
\frac{\gamma}{(\hbar \omega - \hbar \omega_0 - i \gamma)}
\end{equation}
where $g_c$ is the peak-gain coefficient, defined by the inversion of population, $\omega_0$ is the resonance frequency, $c_0$ is the vacuum speed of light, $n_r$ is the refractive index and $\gamma$ is the level broadening.\\
The active region of the THz \ac{QCL} consists of a MBE-grown GaAs/AlGaAs heterostructure, comprising 200 periods of a four-well single-stack hybrid bound-to-continuum/longitudinal-optical phonon depopulation design resulting in a total height of 13.3\,$\mathrm{\mu m}$\,\cite{Amanti_2009}. To balance radiation losses and self-heating, particularly in \ac{cw} operation, a bend radius of 1000\,$\mathrm{\mu m}$ for the inner ring was selected, measured at the inside edge. The width of the lasers are, as previously described, identical at 16\,$\mathrm{\mu m}$. To achieve high mode confinement combined with a low threshold current density and a reduced waveguide width, a double-metal waveguide design is used where the active region is sandwiched between two metal layers (Fig.\,\ref{fig:C1439_rings_paper})\,\cite{Unterrainer_02, Williams_2005}. Consequently, to maximize coupling between the lasers, it is necessary to maximize the interaction length, which leads to a concentric design. To electrically contact the lasers in a \ac{cw} configuration, each laser features four contacting pads (60x60\,$\mathrm{\mu m}$), connected by wire bonds. To minimize their influence on optical properties, the contacting pads are connected by a narrow bridge (50\,$\mathrm{\mu m}$ long, 10\,$\mathrm{\mu m}$ wide), preventing reflection points and mode distortions. Additional small fins (10x10\,$\mathrm{\mu m}$) were added to the sides of the pads to allow a more reliable wire bonding process while maintaining a small size. 
\section{Results}
\begin{figure}[htbp]
\centering\includegraphics[width=13cm]{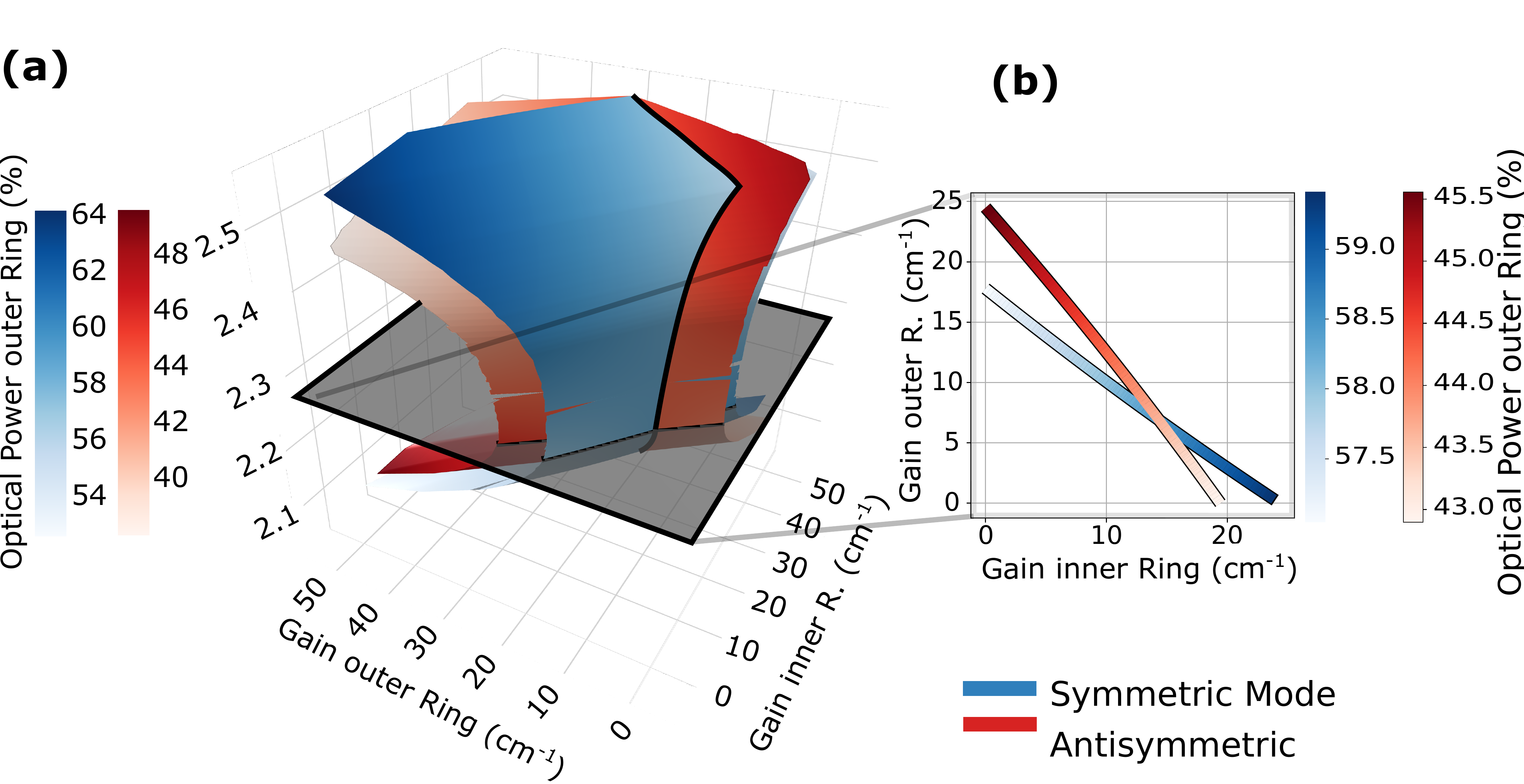}
\caption{Finite element simulation of the concentric ring structure for a transition frequency of 2.28\,THz and 2.4\,THz and a gap size of 4\,$\mathrm{\mu m}$ performed with COMSOL-Multiphysics. (a) Simulation of the threshold gain depending on the frequency detuning and the supermode type. The x- and y-axes denote the gain of the inner and outer ring, and the z axes the frequencies. Depending on the detuning from the transition frequency and the gain, respectively the population inversion, different frequencies and modes reach the threshold condition. The intersection curve between symmetric and antisymmetric supermodes has only low frequency dependence (b) Cutting-plane of the 3D plot at 2.28\,THz showing the threshold condition for the transition at 2.28\,THz. Depending on the detuning from the transition frequency and the gain, different frequencies and supermodes fulfill the threshold condition. The intersection point between the gain curves denotes the point of equal loss in symmetric and antisymmetric supermodes.}
\label{fig:gain_inset}
\end{figure}

\begin{figure}[htbp]
\centering\includegraphics[width=6.5cm]{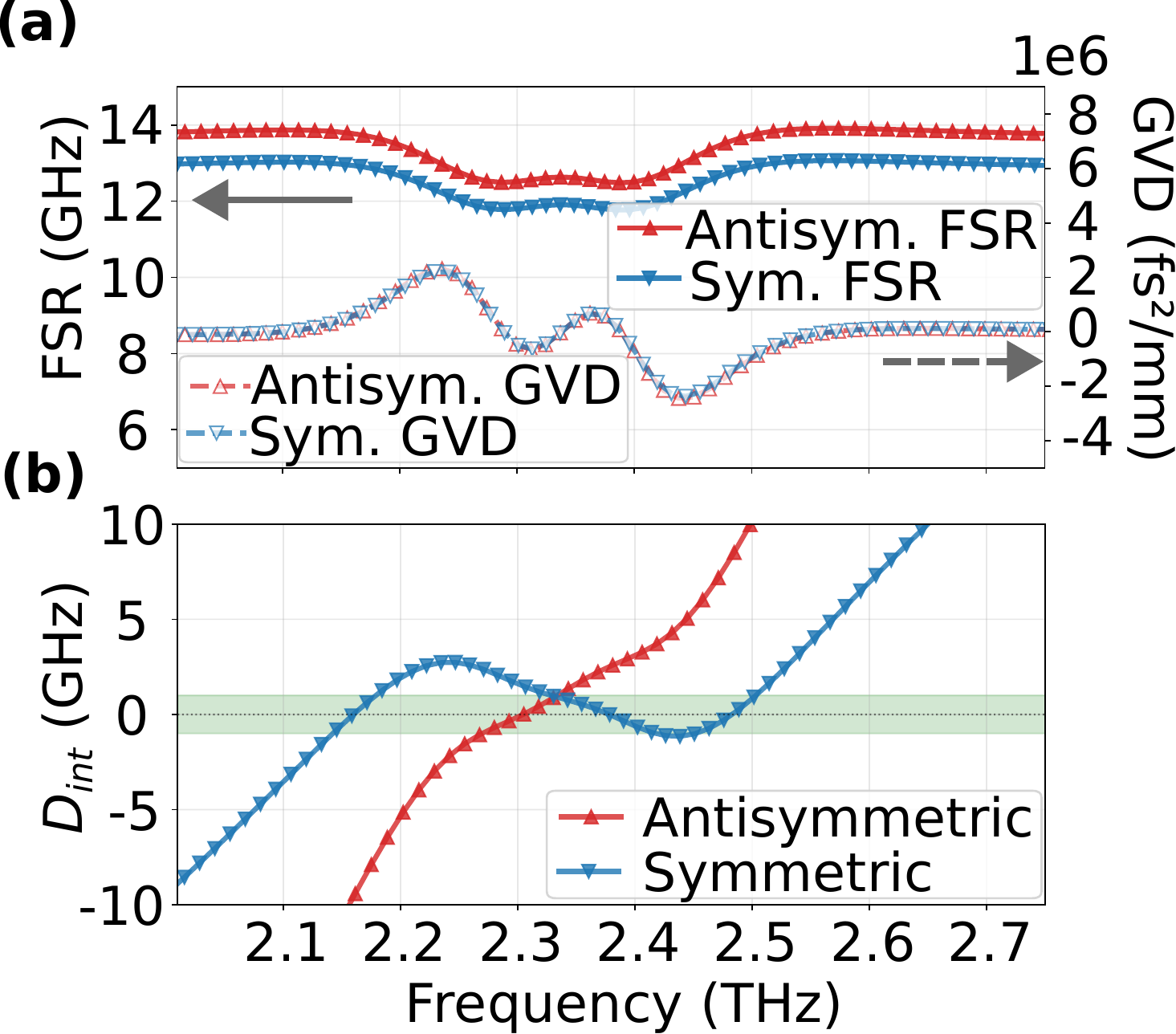}
\caption{(a) \ac{FSR}s of the symmetric and antisymmetric modes, and their respective group velocity dispersions including the influence of the two lasing transitions at frequencies of 2.28\,THz and 2.4\,THz, assuming a gain of 18\,cm$^{-1}$ for the inner laser and 9\,cm$^{-1}$ for the outer laser b) Integrated dispersion, demonstrating the extended locking range (green region) achieved by combining both symmetric and antisymmetric modes.}
\label{fig:integrated_dispersion}
\end{figure}

We fabricated double ring samples featuring gap sizes of 4\,$\mathrm{\mu m}$, 6\,$\mathrm{\mu m}$, 8\,$\mathrm{\mu m}$ and 10\,$\mathrm{\mu m}$. To operate the lasers, we connected both the inner and outer ring-laser to independent power sources, either \ac{cw} or pulsed. Generally, \ac{cw} operation offered a more stable operation due to the constant temperature of the active region, which is favorable for comb operation and beat-note measurements. Nevertheless, operating both lasers at their maximum power in \ac{cw} led to an increase in the heat-sink temperature of about 10\,K, owing to the limited cooling capacity of the cryostat. Consequently, this constrained the performance of the lasers. To mitigate self-heating, we performed pulsed measurements. For this, the lasers were electrically pulsed with a duty cycle of 20\% at 5\,kHz allowing a heat-sink temperature of 5\,K, while maintaining a stable pulse shape and high output power.\\
We initially characterized our rings by biasing them individually while the second ring is unbiased.
Typically, single ring resonators with a comparable width show multiple operation regimes including a comb state\,\cite{Jaidl_21}. In contrast, when biasing a single ring within the concentric-ring structure, we observed only narrow bandwidth multi-mode emission. We attribute this to the high losses in the unbiased ring. In this configuration, the measured threshold current densities of the inner and outer rings are similar, which is supported by the simulation shown in Fig.\,\ref{fig:gain_inset}(b). Given that the threshold gain depends on the total losses, the increased losses due to the passive cavity led to an increased threshold gain compared to a single ring structure. Consequently, as the gap size increases, the dynamic range increases as a result of reduced coupling. Between the smallest and largest fabricated gaps, we observed an increase in the dynamic range by a factor of approximately 3. In pulsed operation, the maximum output is observed for a bias current between 0.21\,$\mathrm{kA/cm^2}$ and 0.26\,$\mathrm{kA/cm^2}$, being higher on the outer ring.\\
In \ac{cw} operation, both the threshold current and maximum operation current increased similarly across all lasers, by approximately 28\% compared to pulsed operation.
The measured optical power of the outer laser was significantly higher compared to that of the inner laser. Since the lasers are edge-emitting, the outer laser is blocking the emission of the inner laser, reducing the light-collection efficiency. This effect is most pronounced for the smallest gap sizes, with an increase in the optical power of up to $\sim$\,4.3 times between the inner and outer laser. Additionally, due to the decreased coupling and, hence, losses in the unbiased ring, the maximum output power increased with increasing gap sizes. The increase was most notable for the inner ring, where the maximum output power increased by a factor of 5.6 by increasing the gap size from 4\,$\mathrm{\mu m}$ to 10\,$\mathrm{\mu m}$.
%
When biasing both rings simultaneously, we observe a remarkable increase in the dynamic range of the bias currents and of the combined optical output power (Fig.\,S2(a)). This effect was most pronounced for a gap of 4\,$\mathrm{\mu m}$, where the dynamic range on the inner ring increased by a factor of approximately 8 and on the inner ring by a factor of approximately 4. The combined maximum output power for this structure, when both lasers are biased simultaneously, is approximately three times higher than the sum of the individually operated lasers. To measure these \ac{LIV} characteristics, both lasers are connected to individual pulse generators, allowing for an independent sweep of the current of both lasers.\\
In a simplified model, below the threshold, the gain of a single \ac{QCL} increases linearly with the applied current. Above the threshold, the gain clamps and further increase in the current leads to an increasing photon flux\,\cite{Faist_13}. In the coupled system, increasing the current (gain) on one ring reduces the threshold gain in the other ring, due to decreasing total losses (Fig.\,\ref{fig:gain_inset}). Consequently, this results in an increase in the total photon flux and consequently higher current due to enhanced photo-induced transport.
 \begin{figure}[htbp]
\centering\includegraphics[width=13cm]{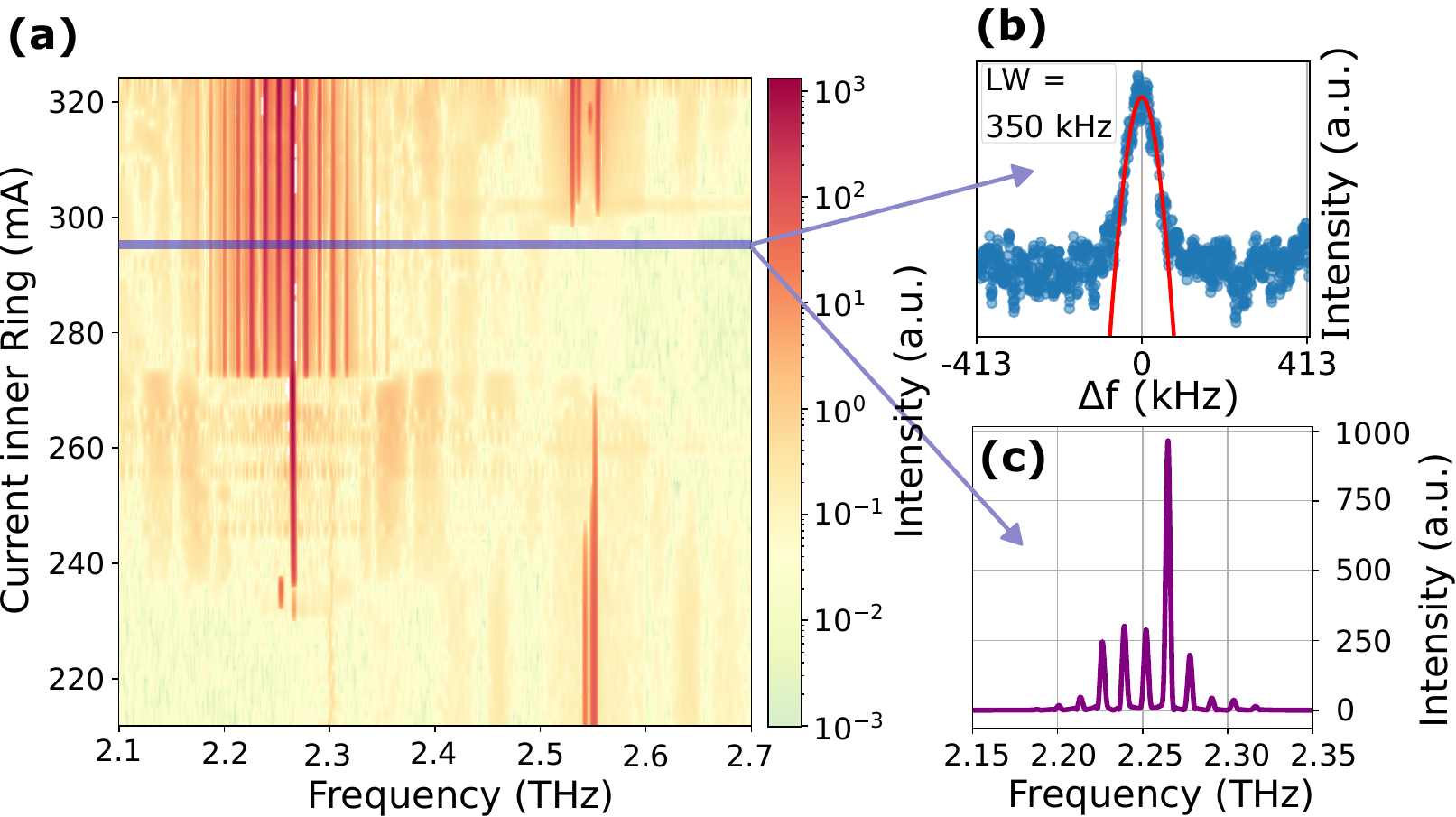}
\caption{(a) Spectra of the coupled ring structure with a gap of 4\,$\mathrm{\mu m}$. The outer ring is biased \ac{cw} at a constant current of 297\,mA, while the bias of the inner ring is swept. At low bias of the inner ring, only two modes at 2.55\,THz are observed until the bias reaches 240\,mA. At this point an additional mode at 2.28\,THz emerges. The comb regime is reached at 274\,mA with up to 15 comb lines, accompanied by depletion of the modes at 2.55\,THz. Further increase of the bias above 300\,mA causes the 2.55\,THz modes to reappear. (b) Optical beat-note measurement at a bias of 296\,mA on the inner ring and 297\,mA on the outer ring with a center frequency of 12.9\,GHz and a fitted linewidth of 350\,kHz. (c) Corresponding optical spectrum under identical conditions to those in (b)}
\label{fig:spectra0918}
\end{figure}
Given the relationship between current, sub-threshold gain, and system losses, the current emerges as the favorable control parameter. To measure the modulation of the optical spectrum, the current on one laser is stabilized while the applied current on the second laser is swept across its full dynamic range. To facilitate more reliable detection of the spectral modulation, it is advantageous to stabilize the current on the outer ring above the threshold due to the higher collection efficiency compared to the inner ring.\\
The fascinating behavior of our dual-ring device can be very impressively followed from the spectral development. In Fig.\,\ref{fig:spectra0918}, the measured \ac{cw} spectra obtained from a dual ring structure with a gap size of 4\,$\mathrm{\mu m}$ are shown. In this measurement, we stabilized the current at the outer ring at its maximum value (obtained for an unbiased inner ring) and swept the current at the inner ring from 212\,mA to 324\,mA. This allows full modulation of the inner ring, while ensuring that the current on the outer ring is below the \ac{NDR} regime. In \ac{cw} operation the threshold current at the inner laser with unbiased outer laser is 287\,mA and the maximum operation current is 296\,mA. At low bias on the inner ring, a dual mode at 2.55\,THz was observed. At 230\,mA, an additional mode at 2.28\,THz emerged, indicating modulation far below the threshold of the inner laser. At a bias of 272\,mA, still below the threshold, pure single-mode operation at 2.28\,THz was achieved. At higher bias, the discrete modes at 2.55\,THz disappeared and the laser entered a broad multi-mode regime, with up to 15 modes between 2.17\,THz and 2.36\,THz. Notably at an increased bias of 300\,mA, these discrete modes reemerged in addition to the multi-mode operation at 2.28\,THz, initially as a single-mode before transitioning to dual-mode, three-mode, and finally four-mode operation (Fig.\,\ref{fig:spectra0918}).\\
The equidistant mode spacing of the multi-mode regime suggests the presence of comb operation. 
To validate this, the optical beat note was measured. For this a zero-bias TeraFET\,\cite{Regensburger_18} and a spectrum analyzer are used, allowing a direct measurement of the optical beating frequency and linewidth. This measurement revealed a narrow and stable beat note with a linewidth of 350\,kHz (Fig.\,\ref{fig:spectra0918}(c)), indicating a fixed phase relation between the modes. The measured linewidth is comparable to previously reported data from single-ring structures \cite{Jaidl_25}. This shows that the dual ring structure allows to obtain comb formation.\\
Injecting an RF signal onto the bias line of one of the two lasers enables locking the comb to the external RF source. This resulted in additional modulation and linewidth reduction of the optical beat note down to the Hz level, limited primarily by the RF equipment. This is essential for high precision measurement applications like dual-comb spectroscopy\,\cite{Newbury_10}.
 \\
To understand the different operation regimes and especially the comb formation in our double-ring system, we performed a finite element simulation using COMSOL Multiphysics (Fig.\,\ref{fig:gain_inset}). The simulation includes the gain dependent complex reflective index of the active region including the free carrier contribution arising from the contact layers as well as the absorption of the top and bottom gold. Due to the symmetries in our system, we employed a two-dimensional rotational symmetric simulation.\\
For each lasing frequency, in principle an antisymmetric and a symmetric supermode are possible. The gain and hence the population inversion required for each supermode to start lasing depends not only on the total gain but also on the gain ratio between inner and outer laser. This dependence arises from the distinct field distributions between the inner and outer rings of these modes, with the symmetric mode being more dominant in the outer ring and the antisymmetric mode in the inner ring. Depending on the gain ratio between inner and outer laser, either the symmetric or the antisymmetric mode is more likely to start lasing, as it requires the lower gain. Figure\,\ref{fig:gain_inset}(b) illustrates this for the transition frequency at 2.28\,THz. If only the outer ring is biased, the symmetric mode is more likely to start lasing, as the required gain is $\sim 30\%$ lower compared to that of the antisymmetric mode. This behavior reverses if the inner ring is biased solely, allowing for switching between symmetric and antisymmetric modes by biasing the inner or the outer ring.\\
The optical power distribution between the inner and outer rings depends not only on the supermode but also on the frequency. In general, the lower gain mode (symmetric or antisymmetric) exhibits an increase in the overlap factor with the outer laser with increasing frequency. Concurrently, the overlap factor with the inner laser decreases.\\
Specifically, when the outer ring is biased dominantly according to the conditions depicted in Fig.\,\ref{fig:gain_inset}(b), the symmetric mode becomes lasing. The overlap factor of the symmetric mode with the biased region (outer ring) increases with increasing optical frequency in contrast to the antisymmetric mode, indicating that higher frequencies are favored under these conditions. Conversely, when the inner ring is biased dominantly, the antisymmetric mode is favored, concentrating most of the optical power within the inner ring. This leads to a preference for lower frequencies. This process is identical for each individual transition. Therefore, it only dominates within a single transition or very close spaced transitions. Additionally, the gain curves of the symmetric and antisymmetric modes can be tuned by changing the width of the lasers and the gap size. The gain curves become increasingly parallel with decreasing width of the air gap, owing to enhanced hybridization. Furthermore, in the case of non-parallel gain curves the position of equal losses in the symmetric and antisymmetric supermodes can be tuned by adjusting the width of the lasers. This explains why a careful tuning of the dimensions of the lasers is necessary, such that the gain crossing occurs at a point reachable by tuning the gain of the two lasers. This is essential to enable switching between symmetric and antisymmetric modes. 
\begin{figure}[h!]
    \centering \includegraphics[width=13cm]{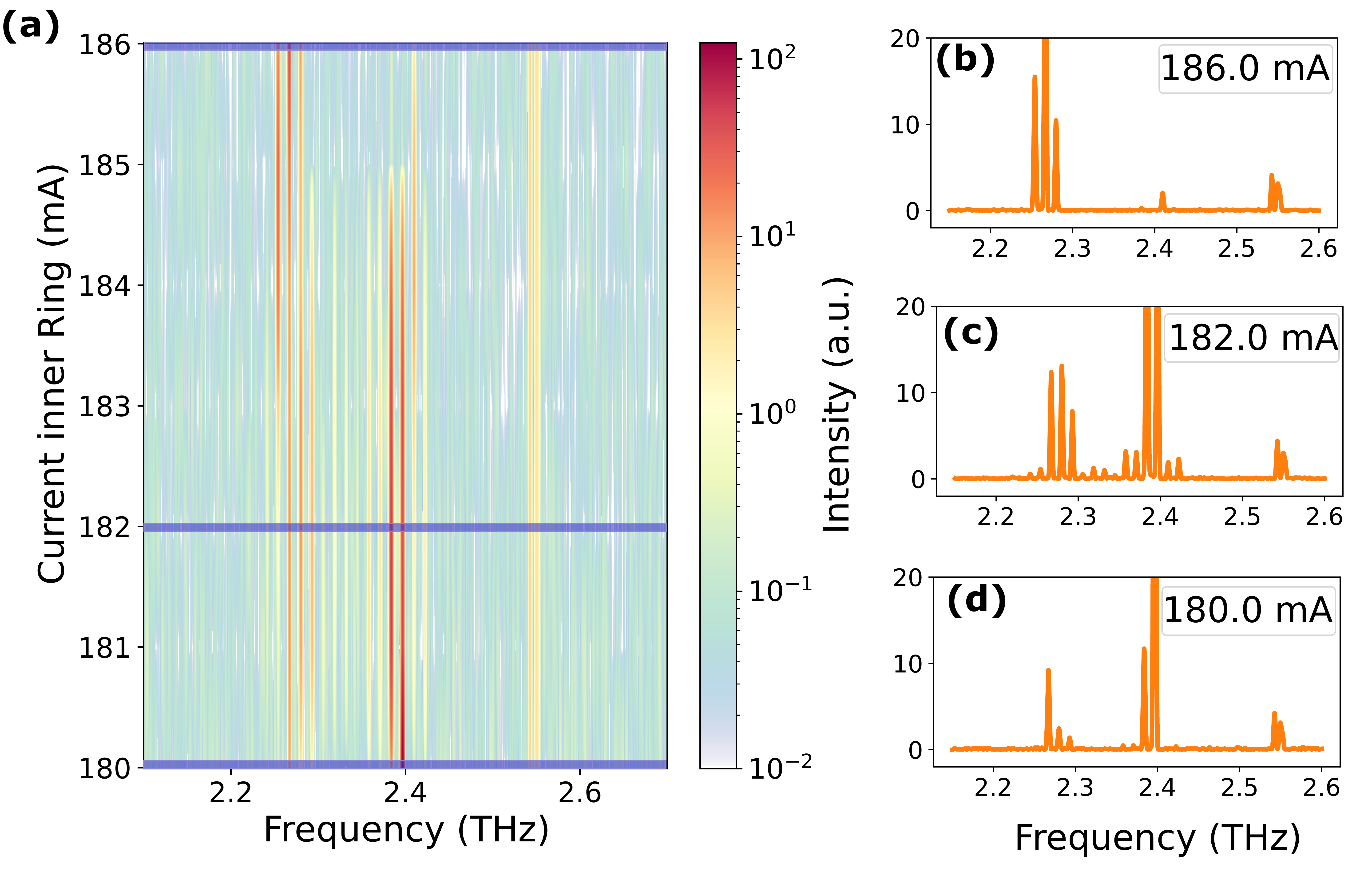}
    \caption{(a) Spectra of a structure with 4\,$\mathrm{\mu m}$ gap size measured at the crossing point between the symmetric and antisymmetric mode. The current of the outer ring is fixed at 246\,mA while the bias on the inner ring is swept. Both lasers are operated in pulsed configuration. (b) The spectra at a current of 180\,mA at the inner laser shows modes at 2.28\,THz and 2.4\,THz as well as 2.55\,THz. (c) At 182\,mA additional modes between 2.28\,THz and 2.4\,THz are appearing gaping the two transitions. (d) At 186\,mA the modes at 2.28\,THz are dominating.
    }
    \label{fig:spectra_3um_pulsed}
\end{figure}
The measured spectra in Fig.\,\ref{fig:spectra_3um_pulsed} closely mirror the behavior predicted by the simulation.
Given that the transitions at 2.28\,THz and 2.4\,THz are very close in frequency, the non-zero gain between these frequencies is bridging them (Fig.\,\ref{fig:gain_inset}(a),\,\ref{fig:integrated_dispersion}(a)). This allows a switching between these two frequencies by switching between symmetric and antisymmetric modes. The measured spectra in Fig.\,\ref{fig:spectra_3um_pulsed}, measured in pulsed operation demonstrates this behavior. By increasing the current on the inner laser while the current on the outer laser is fixed at the maximum current in individual operation (246\,mA\,$\simeq$\,0.238\,\,$\mathrm{kA/cm^2}$), we observe a transition from the higher frequency optical transition, measured at a center frequency of $\sim 2.4$\,THz to the lower frequency transition, measured at $\sim 2.28$\,THz. Additionally, we observe a spectral broadening of the comb. This occurred only within an extremely narrow bias window, of approximately 1\% of the total tuning range. This broadening was observed only in pulsed operation, which we attribute to the lower temperature of the active region, resulting in lower losses.
According to our simulations in Fig.\,\ref{fig:gain_inset}(a), if the outer laser is above threshold while the gain on the inner laser is increased, the threshold gain on the outer laser follows two distinct curves for the symmetric and the antisymmetric mode. Due to the non-parallel nature of these curves, a point of equal losses in both supermodes exists. This transition point occurs at a gain of approximately 18\,$\mathrm{cm^{-1}}$ at the inner ring and 9\,$\mathrm{cm^{-1}}$ at the outer ring and exhibits only low frequency dependence.\\
The dominating process of frequency comb generation in \ac{QCL}s is the high third order nonlinearity ($\chi^{(3)}$) and the resulting \ac{FWM} \cite{Burghoff_14, Friedli_13}. In the case of our coupled laser system the \ac{FWM} process is not limited to a single supermode type. The high hybridization of the resonator modes allows for an energy pathway between the different supermodes facilitated by \ac{FWM}. This enables a low-dispersion path between the supermodes, resulting from the generally lower \ac{FSR} of the symmetric modes compared to the antisymmetric modes. Consequently, this results in a reduced respective group velocity dispersion and total integrated dispersion, thereby enhancing the locking range and broadening the comb spectrum (Fig.\,\ref{fig:integrated_dispersion}). This
effect was previously explored by Tikan et al.\,\cite{Tikan_2021}.
\begin{figure}[htbp]
\centering\includegraphics[width=11cm]{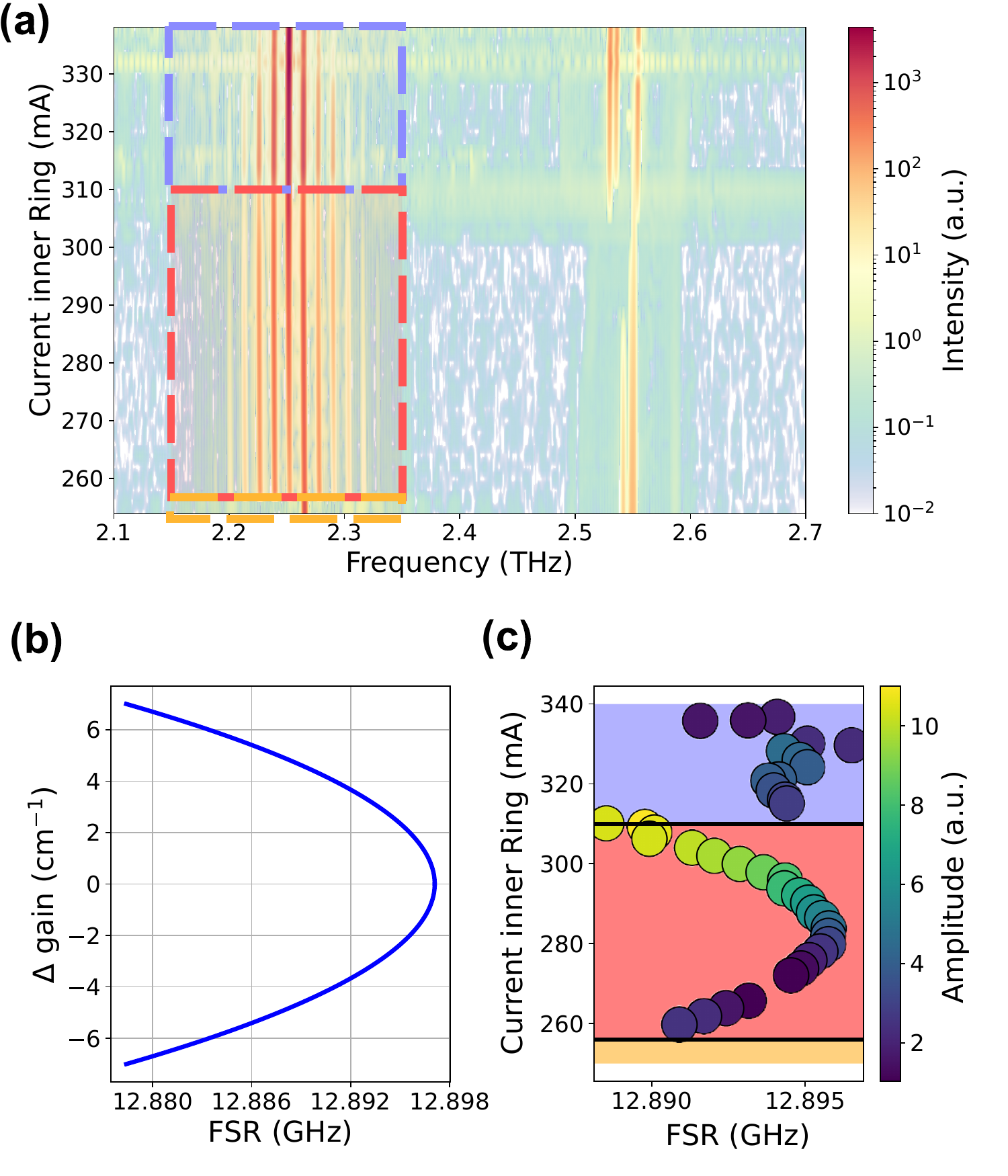}
\caption{(a) Spectra of a structure with a gap size of 4\,$\mathrm{\mu m}$ measured at a fixed bias of 315\,mA on the outer ring. Both lasers are driven in \ac{cw}. At low bias no comb but a single-mode is apparent (orange) at increasing bias the comb regime is reached (red). The \ac{FSR} in this region can be tuned by changing the bias. (b) The beat-note frequency of the antisymmetric mode was calculated using the Hamiltonian matrix formalism in Eq.\,\eqref{eq:hemiltonian_2x2_FSR} with a coupling strength-to-detuning ratio of 3.5, corresponding to a resonator distance of 4\,$\mathrm{\mu m}$. (c) Measurement of the electric beat-note frequency. The maximum tuning range is about 7\,MHz according to (b). This corresponds to a a maximum gain detuning of about $\pm$5\,$\mathrm{cm^{-1}}$.}
\label{fig:beatnote_comb}
\end{figure}
To prove that these lower frequencies modes at $\sim 2.28$\,THz indeed correspond to antisymmetric modes, a \ac{cw} measurement was performed allowing the measurement of the electrical beat-note frequency (Fig.\,\ref{fig:beatnote_comb}). For this measurement the outer laser was fixed at its maximum operation current similar to the pulsed measurement. Due to the higher temperature of the active region in \ac{cw} operation, the maximum current of the outer ring increased to 315\,mA (0.304\,$\mathrm{kA/cm^2}$). The optical spectrum features a transition from single-mode to comb operation and the measured electrical beat note follows a near perfect arc shape (Fig.\,\ref{fig:beatnote_comb}(c)).\\
The splitting between symmetric and antisymmetric modes depends on the ratio of coupling strength to detuning, which is fixed by the laser design and the gain detuning ($\Delta g$). According to the matrix formalism in Eq.\,\eqref{eq:hemiltonian_2x2_FSR} and Fig.\,\ref{fig:hybridisation} the optical power distribution between the rings and the \ac{FSR} splitting depend on the gain detuning between the inner and the outer laser. This dependence is maximized when the ratio of the coupling strength to the detuning is approximately unity. \\
When the coupling strength significantly exceeds the detuning, the mode splitting becomes independent of the gain ratio and is dictated by a power law. Conversely, at low coupling, the splitting is dominated by the length difference. Consequently, to allow a tuning of the \ac{FSR} by the gain detuning, the ratio between coupling strength and detuning must be close to the ideal value of 2.\\
In our system, coupling strength and detuning are linked by the gap size and cannot be tuned independently. At a gap size of 4\,$\mathrm{\mu m}$ the simulation shows that the coupling-to-detuning ratio is approximately 4, which allows for sufficient tuning while securing a high degree of hybridization (Fig.\,\ref{fig:hybridisation}).\\
Consequently, given a fixed coupling-to-detuning ratio, increasing the gain detuning results in an increase in the \ac{FSR} of the symmetric mode and a decrease of the antisymmetric mode. Essentially, increasing the gain/loss difference reduces the splitting (Fig.\,\ref{fig:hybridisation}). As a result, the beat-note frequency of the antisymmetric modes follows a parabolic curve with a maximum frequency at zero gain detuning, while symmetric modes exhibit a minimum beat-note frequency at zero gain detuning.\\
The measured electrical beat note in Fig.\,\ref{fig:beatnote_comb}(c) matches the calculated arc shape of an antisymmetric mode with a gap size of 4\,$\mathrm{\mu m}$ near perfect (Fig.\,\ref{fig:beatnote_comb}(b)), suggesting a maximum gain detuning of $\pm 5\,\mathrm{cm^{-1}}$. This proves that the observed mode is indeed antisymmetric.
\section{Conclusion}
In this work, we have demonstrated a novel approach for electrically tuning optical frequency combs in coupled quantum cascade ring lasers. By exploiting the symmetry breaking between symmetric and antisymmetric supermodes, we achieve high-speed electro-optic tuning of the emission spectra by utilizing the outer ring laser as a master, while the inner ring serves as an integrated modulator. This allows us to switch between different operation spectral regions ranging from 2.28\,THz to 2.55\,THz as well as switching between a comb state with up to 11 modes and single-mode operation. The matrix formalism employed accurately predicts the frequency splitting behavior and modal dynamics under various coupling and detuning conditions, further validated by comprehensive numerical simulations, which allowed for the optimization of the resonator design. Moreover, the observed spectral tunings were attributed to active modulation of the gain/loss ratio within the coupled-cavity system, rather than independent lasing from the modulator, as modulation occurs well below the inner rings lasing threshold. Our results pave the way for further investigations into enhancing control of optical frequency combs and exploring additional functionalities enabled by symmetry-control. Our findings highlight the potential of this technique for integrated spectroscopy and sensing applications, offering flexible electrical tunability without the need for complex and expensive RF equipment.
\\
\noindent
\begin{backmatter}
\bmsection{Acknowledgment} The authors acknowledge financial support from the Austrian Science Fund FWF (TeraLearn
P35932-N, Everest I 5772) and funding through the Hochschuljubiläumsstiftung der Stadt Wien (H-496620/2024). A.M.A.
acknowledges the support of the United States Air Force Office of Scientific Research (FA8655-23-1-7070, FA8655-25-1-7004).
\bmsection{Disclosures} The authors declare no conflicts of interest.
\bmsection{Data availability} Data underlying the results presented in this paper are not publicly available at this time but may be obtained from the authors upon reasonable request.
\end{backmatter}

\bibliography{bibliography}
\end{document}